\documentclass[11pt]{article}
\usepackage{a4wide,epsfig}
\def\sd{\strut\displaystyle}

\begin{document}
\vspace{3cm}
\parindent 1.3cm
\title{\bf $\bf \eta$ decays involving photons\footnote{Contribution to
the Eta Physics Handbook, Workshop on Eta Physics, Uppsala 2001.}}
\author{Ll. Ametller\\
\ \\ 
Departament F{\'\i}sica i Enginyeria Nuclear\\
Universitat Polit\`ecnica de Catalunya, E-08034 Barcelona, Spain\\  
}
\date{\ }
\maketitle

\vspace{2cm}
\begin{abstract}
The decays of the $\eta$ meson are reviewed in
the framework of Chiral Perturbation Theory. Particular attention is
devoted to the electromagnetic channels involving photons, either in
the final or intermediate states. 
\end{abstract}

\vspace{5cm}

\begin{center}
$\begin{array}{l}
\hbox{\rm {\bf PACS:} 12.39.Fe, 13.40.Hq, 14.40.Aq, 12.40.Vv }\\
\hbox{\rm {\bf Keywords:} Chiral Lagrangians, Electromagnetic Decays,} \\
\hbox{\rm ~~~~~~~~~~~~~~~~~~$\pi, K$ and $\eta$ Mesons, Vector-meson Dominance.}
\end{array}$
\end{center}

\thispagestyle{empty}
\newpage\setcounter{page}{1}

Several experimental setups have been recently 
proposed to study the physics of
the $\eta$ meson. Some are already operative, like the SND detector 
at the VEPP-2M collider in Novosibirsk
\cite{SND} and KLOE at the Daphne collider in Frascati \cite{Daphne}. 
Together with the Wasa detector at the CELSIUS storage ring in Uppsala, 
which should start taking data in early 2002, 
they will hopefully collect a copious 
number of $\eta$ mesons. The existence of these facilities brings the 
possibility to study many of its decays. 
Most of the $\eta$ decays detected so far \cite{PDG}, like 
$\eta\to\gamma\gamma$, $\eta\to
\pi^+\pi^-\gamma$ or $\eta\to\pi^0\gamma\gamma$,
involve external photon(s) and thus proceed unambiguously through
electromagnetic (e.m.) interactions. Some channels into leptons,
$\eta\to \mu^+\mu^-$, for instance, proceed through photonic intermediate states
and share the e.m. origin too.
The remaining ones, $\eta\to \pi^+ \pi^- \pi^0$ and $\eta\to 3\pi^0$,
are purely hadronic but
violate G--parity (or isospin symmetry), and have therefore small branching ratios,
at the level of the e.m. ones. 
We present here a survey  on theoretical predictions for $\eta$ decays, focused
mainly on those involving photons.
A complete discussion on  $\eta\to 3 \pi$ can be found in this Handbook \cite{gasser,hans,holstein}.
In addition, weak decays, with emphasis on $C$ and/or $CP$ violation, are discussed in
Ref.\cite{Shabalin}.

Being the $\eta$ mass just above the kaon mass, $m_\eta\simeq 550~MeV$, the
variety of e.m. $\eta$ decays represents, in principle, an excellent 
laboratory to test
Chiral Perturbation Theory (ChPT) \cite{GASSER}, both in the odd  and
the even intrinsic parity sectors.
However, two main drawbacks must be mentioned. One is 
the $\eta-\eta'$ mixing which requires the inclusion of the non Goldstone boson
$\eta_1$ (a pure and massive $SU(3)$ singlet)
in the ChPT framework. This has been discussed in Ref.\cite{taron} from
the theoretical point of view and no definite predictions can be done since, 
at one-loop, there are many unknown free parameters. 
The second difficulty is the presence of counter-terms at
next-to-leading ChPT order.
Most of the existing theoretical predictions 
assume resonance saturation of the counter-terms. Finally, for some
processes, definite ChPT predictions are not available and we present
instead estimates based on phenomenological Vector Meson models.

We give in Section $1$ a brief review on the treatment of the
$\eta$--particle in ChPT. Section $2$ is devoted to the odd intrinsic
parity $\eta$  decays (i.e.
those of the anomalous sector, proceeding through the Wess Zumino
Lagrangian),
whereas the even-parity ones are presented in Section $3$.
One multi-body odd-parity final state, 
$\eta\to\pi\pi\gamma\gamma$, is strongly related to $\eta\to\pi\pi\pi$ 
and its discussion is relegated to Section $4$.

\section{$\bf \eta$--particle in Chiral Perturbation Theory}

ChPT is discussed in detail by Gasser \cite{gasser} and Bijnens \cite{hans} in
this Handbook. Here we give a brief theoretical description 
in order to fix our notation and for self-consistency. 

ChPT provides an accurate description of the 
strong and electroweak interactions of pseudoscalar mesons $P$ at low 
energies \cite{GASSER}. In this sense and at lowest order, ChPT is 
essentially equivalent to Current Algebra (CA) \cite{DASHEN}. At higher orders,
loop effects appear giving rise to both finite corrections and 
divergences. The former are known to improve the lowest order (or CA) 
predictions, while the latter require the introduction of sets of counter-terms
--whose number depends on the perturbative order one is working-- and restore  
the renormalizability of the theory, order by order in the perturbative expansion.

The nonet of pseudoscalar mesons $P$ is described in terms
of the $SU(3)$ octet and singlet matrices \cite{hansanom}
\begin{equation}
  \label{M_8}
P_8=\pmatrix{ 
{\sd\pi^0 \over \sd\sqrt 2}+{\sd\eta_8 \over \sd\sqrt{6}} &  \pi^+  &  K^+    \cr
\pi^-   & -{\sd\pi^0 \over \sd\sqrt 2} + {\sd\eta_8 \over \sd\sqrt{6}} &  K^0 \cr
 K^-    &     \bar K^0      & {-\sd 2\over \sd\sqrt{6} } \eta_8  \cr }, 
\qquad P_1={\sd{1\over \sqrt{3}}}\eta_1 \hbox{I,}
\end{equation}
which appear in the ChPT Lagrangian through the parametrization
\begin{equation}
\label{SIGMA}
\Sigma\equiv\Sigma_8\Sigma_1=\Sigma_1\Sigma_8=\exp\left(\sd{2i\over 
f}(P_8+P_1)\right),
\end{equation}
with $f=130.7~\hbox{MeV}$ \footnote{We use the  PDG's normalization 
\cite{PDG} for the pion decay constant. Notice that in most theoretical papers,
$F = f/\sqrt{2}$ is used.},
as can easily be deduced from the charged pion 
decay, with e.m. corrections included \cite{GASSER,PDG}. 
The transformation properties of $\Sigma$ under 
chiral $U(3)_L\times U(3)_R$ are given by $\Sigma \to U_L\Sigma 
U_R^\dagger$ and the physical $\eta$ and  $\eta'$  particles deviate from being a pure
$SU(3)$--octet and  $SU(3)$--singlet, respectively. The $\eta-\eta'$ mixing 
is conventionally formulated in terms of $\theta$ and/or $\phi=\theta +
\hbox{\rm arctan} \sqrt 2$, which refer to the octet-singlet and non-strange-strange 
basis, respectively,
\begin{eqnarray}
\label{MIXING}
\eta&=&\cos\theta~\eta_8 -\sin \theta~\eta_1 = 
\cos \phi ~(u\bar u + d \bar d)/\sqrt 2 - \sin\phi ~s \bar s \nonumber\\
\eta'&=&\sin\theta~\eta_8 +\cos \theta~\eta_1=
\sin \phi ~(u\bar u+ d \bar d)/\sqrt 2 + \cos\phi ~s \bar s 
\end{eqnarray}
This formulation of the 
$\eta-\eta'$ states mixing has been mainly used in the analysis of radiative
transitions  \cite{Donoghue,BIJNENS1,BIJNENS2}. 
For even intrinsic parity vertices with $4$ pseudoscalars, our treatment of the
$\eta-\eta'$ mixing will follow the standard ChPT treatment, i.e., 
the $\eta$ singlet effects are 
included in the $O(p^6)$ counter-term $L_7^r$ (see below).

We assume that nonet symmetry gives a good description of the whole nonet
except for the singlet mass which gets an extra term, correcting the $U_A(1)$
problem. Because of its heavy mass we neglect the singlet component in
loops and use the octet component with the mass according to the
Gell-Mann-Okubo formula. The mixing angle (\ref{MIXING}) is not negligible
and can have rather large effects in $\eta$ decays. 
This treatment also fits the mass matrix
reasonably well if the loop corrections to the masses are taken into
account \cite{GASSER}.

The lowest order Lagrangian of ChPT ---order two in particle four--momenta 
or masses, $O(p^2)$, and ignoring the extra singlet mass term--- is
\begin{equation}
\label{L2lluis}
{\cal L}_2=\sd{f^2\over 8} tr (D_\mu\Sigma D^\mu\Sigma^\dagger + \chi 
\Sigma^\dagger +\chi^\dagger \Sigma),
\end{equation}
where the covariant derivative
\begin{equation}
\label{COVDER}
D_\mu\equiv \partial_\mu + i e A_\mu [Q,\Sigma]
\end{equation}
contains the photon field $A_\mu$ and the quark charge matrix $Q$  
[$Q=\hbox{diag}(2/3,-1/3,-1/3)$] thus generating the e.m.
couplings of charged P's 
through the commutator $[Q,\Sigma]$. The non--derivative terms in Eq. 
(\ref{L2lluis}) with $\chi=\chi^\dagger = B~\cal M$ contain the quark mass matrix 
$\cal M$ 
[${\cal M}=\hbox{diag}(m_u,m_d,m_s)$] and lead to
\begin{equation}
\label{B}
\sd{{1\over 2}B={m_K^2\over m_u+m_s}= {m_\pi^2\over m_u+m_d}= 
\frac{\sqrt{3} m^2_{\pi^0\eta}}{m_u-m_d}=
{\Delta 
m_K^2\over m_d-m_u}},
\end{equation}
where $m^2_{\pi^0\eta}$ is the mass matrix element
describing $\pi^0$--$\eta$
mixing and
\begin{equation}
\label{TADPOLE}
\Delta m_K^2\equiv (m_{K^0}^2-m_{K^+}^2)_{QCD} = 6.0\times 
10^{-3}~\hbox{GeV}^2
\end{equation}
is the kaon mass squared difference (e.m. self energies removed).
Its numerical value is an average between 
the result \cite{GASSER}
$\Delta m_K^2= (m_{K^0}^2-m_{K^+}^2 - m_{\pi^0}^2 + m_{\pi^+}^2) = 5.3\times
10^{-3}~\hbox{GeV}^2,$
\noindent
following from Dashen's theorem \cite{DASHEN}, and independent estimates 
\cite{BRAMON} 
(including improved versions of Dashen's theorem \cite{DTnew}) 
leading to $\Delta m_K^2$ in the range $(6.5 - 7.0)\times 
10^{-3}~\hbox{GeV}^2$. 

The next order Lagrangian, $O(p^4)$, contains a series of ten 
counter-terms identified and studied by Gasser and Leutwyler \cite{GASSER},
\begin{equation}
\label{L4}
\begin{array}{l}
{\cal  L}_4= L_1(tr D_\mu\Sigma D^\mu\Sigma^\dagger)^2
+ L_2 (tr D_\mu\Sigma D_\nu\Sigma^\dagger)^2
+ L_3 tr (D_\mu\Sigma D^\mu\Sigma^\dagger D_\nu\Sigma D^\nu\Sigma^\dagger)\\
\ \\
+ L_4 tr (D_\mu\Sigma D^\mu\Sigma^\dagger) tr (\chi\Sigma^\dagger+ \Sigma
\chi^\dagger) 
+ L_5 tr \left[(D_\mu\Sigma D^\mu\Sigma^\dagger) (\chi\Sigma^\dagger+ \Sigma
\chi^\dagger)\right] \\
\ \\
+ L_6\left[tr(\chi\Sigma^\dagger+ \Sigma\chi^\dagger)\right]^2 
+ L_7\left[tr(\chi\Sigma^\dagger- \Sigma\chi^\dagger)\right]^2 
+ L_ 8 tr (\chi\Sigma^\dagger \chi\Sigma^\dagger+
          \Sigma\chi^\dagger \Sigma\chi^\dagger) \\
	  \ \\
- i e L_9 F_{\mu\nu} tr (Q D^\mu\Sigma D^\nu\Sigma^\dagger 
                        +Q D^\mu\Sigma^\dagger D^\nu\Sigma)
+ e^2 L_{10} F_{\mu\nu}F^{\mu\nu} tr (\Sigma^\dagger Q \Sigma Q)
\end{array}
\end{equation}
and the Wess--Zumino (WZ) term \cite{WESS}, responsible for the anomaly
and odd intrinsic parity 
processes, whose action (ignoring external fields other than photons) is
\cite{BIJNENS1,Hans} 
\begin{equation}
\label{WZ}
\begin{array}{l}
S_{WZ}= \sd\frac{-i N_c}{240 \pi^2} \int d\Sigma^{ijklm} tr (\Sigma
\partial_i \Sigma^\dagger \Sigma \partial_j\Sigma^\dagger\Sigma \partial_k
\Sigma^\dagger\Sigma\partial_l \Sigma^\dagger\Sigma\partial_m \Sigma^\dagger)\\
\ \\
\ \ \ \ \ \ \ -\sd\frac{e N_c}{48 \pi^2} \epsilon^{\mu\nu\alpha\beta} A_\mu 
tr \left[Q(\partial_\nu\Sigma\partial_\alpha \Sigma^\dagger \partial_\beta 
\Sigma\Sigma^\dagger-
\partial_\nu\Sigma^\dagger\partial_\alpha\Sigma 
\partial_\beta\Sigma^\dagger\Sigma)\right]\\
 \ \\
\ \ \ \ \ \ \ 
-\sd\frac{i N_c e^2}{24 \pi^2}\epsilon^{\mu\nu\alpha\beta} (\partial_\mu A_\nu)
A_\alpha tr\Big(Q^2 \partial_\beta \Sigma\Sigma^\dagger
+ Q^2 \Sigma^\dagger\partial_\beta\Sigma \\
\ \\
\ \ \ \ \ \ \
+\frac{1}{2}Q\Sigma Q \Sigma^\dagger
\partial_\beta\Sigma\Sigma^\dagger
-\frac{1}{2}Q\Sigma^\dagger Q \Sigma
\partial_\beta\Sigma^\dagger\Sigma\Big).
\end{array}
\end{equation}
In the above equations, $F_{\mu\nu}$ is the e.m. tensor
and $N_c$ the number of colors.

Accurate numerical predictions require the knowledge of the counter-term
values. One can rely on experimental data to fix the finite part of
their coefficients, as
was originally done by Gasser and Leutwyler \cite{GASSER} or,
alternatively, saturate them by the contributions of low--lying resonances
(vector mesons usually providing the largest contributions)
at $\mu\simeq M_\rho$ \cite{ECKER}. Vector mesons can be introduced using 
a hidden symmetry formulation \cite{Bando} and assuming ideal 
mixing and nonet symmetry. The vector meson nonet can be described by 
$\rho_\mu=\rho_\mu^a \lambda^a/\sqrt{2} + \rho_\mu^1/\sqrt{3}$. The
relevant Lagrangian with vector mesons at $O(p^6)$ is
\begin{equation}
\label{Bandoc}
\begin{array}{l}
{\cal L}_\rho^\epsilon=\epsilon^{\mu\nu\alpha\beta}\Big(tr\big[a_1 a_\mu a _\nu
a_\alpha \sigma^\dagger D_\beta \sigma + i g a_2 \sigma^\dagger
\rho_{\mu\nu} \sigma\{a_\alpha, \sigma^\dagger D_\beta \sigma\} \\
\ \\
\ \ \ \ \ \ \ + i e a_3
F_{\mu\nu}(\xi^\dagger Q \xi + \xi Q \xi^\dagger)\{a_\alpha,
\sigma^\dagger D_\beta \sigma\}\big]\Big)
\end{array}
\end{equation}
where $\sigma$ is a $3\times 3$ unitary matrix parametrized in terms of a
set of unphysical scalar (compensating) fields that, when gauged away,
generates a mass term for the vector mesons. The Lagrangian in  
Eq.(\ref{Bandoc}) also contains the
following hermitian operators
\begin{equation}
\begin{array}{l}
i \sigma^\dagger D_\mu \sigma= i \sigma^\dagger \partial_\mu \sigma - g
\sigma^\dagger \rho_\mu \sigma + v_\mu, \qquad \quad 
v_\mu=\sd\frac{1}{2i}[\xi(\partial_\mu+ i e Q A_\mu) \xi^\dagger +
\xi^\dagger (\partial_\mu + i e Q A_\mu) \xi], \\
\ \\
\rho_{\mu\nu}= \partial_\mu
\rho_\nu - \partial_\nu \rho_\mu + i g [\rho_\mu,\rho_\nu],
\qquad \qquad
a_\mu=\sd\frac{1}{2i}[\xi(\partial_\mu+ i e Q A_\mu) \xi^\dagger -
\xi^\dagger (\partial_\mu + i e Q A_\mu) \xi ],
\end{array}
\end{equation}   
where $\xi=\exp[(i/f) P]$. In order to identify the contribution of the
vector mesons to the finite part of the renormalized amplitudes one has to
integrate out the vector nonet fields  $\rho$  assuming their masses to be much
larger than all the momenta. Explicit expressions can be found in Ref.
\cite{BIJNENS1}.

\section{Odd intrinsic parity $\bf \eta$ decays}

In this section we discuss the most important odd intrinsic parity 
decays (sometimes called anomalous decays) of the
$\eta$ meson of e.m. origin. 
We start with  $\bf \eta \to \gamma \gamma$,
which gives information on the $\eta-\eta'$ mixing angle, and the
closely related $\eta \to \gamma l^+ l^- $ and $\eta\to\mu^+\mu^-$,
involving leptons in the final state. We consider also the process
$\eta\to \pi^+\pi^-\gamma$ and postpone the discussion of the multi-body
channel $\eta\to\pi\pi\gamma\gamma$ to the
Section $4$.

\subsection{$\bf \eta \to \gamma \gamma$ and $\bf \gamma \gamma^* $}
\label{PGG}

Theoretically, $P\gamma\gamma$ transitions involving on--mass--shell
photons, $k^2=k^{*2}=0$, contain valuable information on the mixing
(quark--content) of the $\eta, \eta'$ mesons. The amplitudes
for $\pi^0,\eta_{8,1} \to \gamma\gamma$ in ChPT are
\begin{equation}
\label{pgg}
A(P\to \gamma\gamma^*)= \frac{-\sqrt{2} C_P\, \alpha}{\pi f_P}
\epsilon^{\mu \nu \alpha \beta} \epsilon_\mu k_\nu \epsilon^*_\alpha
k^*_\beta
\end{equation}
where $C_{\pi^0}=1$, $C_{\eta_8}=1/\sqrt{3}$ and $C_{\eta_1}=2\sqrt{2}/\sqrt{3}$
at lowest order 
and
$\epsilon$ and $\epsilon^*$ are the polarization vectors of the
photons with momenta $k$ and $k^*$, respectively.

The present experimental values are \cite{PDG}
\begin{eqnarray}
\label{val}
 \Gamma(\pi^0  \to \gamma\gamma)&=& 7.74 \pm 0.56~eV \nonumber\\
 \Gamma(\eta  \to \gamma\gamma) &=& 0.46 \pm 0.04~keV \\
 \Gamma(\eta '\to \gamma\gamma) &=& 4.29 \pm 0.15~keV. \nonumber
\end{eqnarray}
Notice that the error for $\eta\to\gamma\gamma$ is the largest one, 
about $10\%$,
considerably bigger than the $\sim 3\%$ error of $\eta'\to\gamma\gamma$.
In fact, the value quoted in (\ref{val}) for the $\eta\to\gamma\gamma$ 
decay width is the average between two types of measurements which are not
in agreement.
Indeed, photon-photon  production measurements
obtain $0.510\pm0.026$~keV, whereas Primakoff production
gets $0.324\pm0.046$~keV. A better measurement of this partial decay
width is
thus clearly needed. This would allow to know the total $\eta$ decay 
width with better precision, since it is computed from the
partial decay rate $\Gamma(\eta  \to \gamma\gamma) $ divided by the
fitted branching ratio for that mode.

It is easy to check that Eqs.(\ref{pgg}) lead to an unambiguous 
prediction for the $\pi^0\to\gamma\gamma$
width of $7.74~eV$, in complete agreement with data.
The situation is different for the $\eta$ and $\eta'$ two photon decays. 
We have to remember that $\eta$ and $\eta'$ consist of a mixture of the
$\eta_8$ and $\eta_1$, with a mixing angle $\theta$, as 
quoted in Eq.(\ref{MIXING}).
Pioneering works in ChPT proceed by
fixing $f_{\eta_8}$ to the value obtained at one loop, 
and used the $\eta$ and $\eta'$ decay
widths to fit $\theta$ and $f_1$. They obtained 
$f_{\eta_8}\simeq 1.3 f_\pi$, $f_{\eta_1}\simeq 1.1 f_\pi$ and $\theta
\simeq -19.5^\circ$
\cite{BIJNENS1}. Most of the existing results for radiative 
$\eta$ decays make use of the above values for the mixing angle and 
decay constants.

Notice however that the description of the $\eta-\eta'$ system in terms
of a universal octet-singlet angle $\theta$ has been shown to be
insufficient when a similar pattern is assumed for the decay constants.
(See Kroll's contribution to this workshop \cite{kroll}.)
Defining decay constants through matrix elements of octet and singlet
axial-vector currents, one is lead to  a description involving 
two decay constants, $f_{\eta_8}$ and $f_{\eta_1}$, and two angles
$\theta_8$ and $\theta_1$ \cite{Leutwyler}. They are related through
\begin{equation}
\pmatrix{
f_\eta^8  &  f_\eta^1 \cr
f_{\eta'}^8  &  f_{\eta'}^1 }  = 
\pmatrix{f_8 \cos~\theta_8 &  -f_1 \sin ~\theta_1\cr
f_8 \sin~\theta_8 &   f_1 \cos ~\theta_1
}.
\end{equation} Alternatively, one can define decay constants
related to axial-vector currents with non-strange and strange quark
content. In this context, one writes
\begin{equation}
\pmatrix{
f_\eta^q  &  f_\eta^s \cr
f_{\eta'}^q  &  f_{\eta'}^s }  = 
\pmatrix{f_q \cos~\phi_q &  -f_s \sin ~\phi_s\cr
f_q \sin~\phi_q &   f_s \cos ~\phi_s
}.
\end{equation}
Recent phenomenological analysis, including new experimental data for the
$\eta\gamma$ and $\eta' \gamma$ form factors \cite{cleo} and
data on the $J/\psi\to\eta'\gamma$ and $J/\psi\to\eta\gamma$ 
transitions, have performed an   
overall fit to data  obtaining \cite{kroll}
$ \theta_8=-22.2^\circ,  \theta_1\simeq -9.1^\circ, 
 f_8 = 1.28 f_\pi,  f_1 \simeq 1.20 f_\pi$,
in reasonable agreement with  the theoretical ChPT corresponding results 
\cite{Leutwyler}
$ \theta_8=-20.5^\circ,  \theta_1\simeq -4^\circ, 
 f_8 = 1.28 f_\pi,  f_1 \simeq 1.25 f_\pi$.
Similarly, using the non-strange strange basis, 
$f_q = (1.07 \pm 0.02) f_\pi, f_s = (1.34\pm 0.06) f_\pi, 
\phi_q \simeq \phi_s \simeq  (39.3\pm  1)^\circ $.
Notice that the fact that $\phi_q \simeq \phi_s$ allows to use a unique
angle $\phi \simeq 40^\circ $ in the non-strange strange basis, 
which is in agreement with
phenomenological analysis \cite{krolls} and also with a recent
determination at Daphne \cite{ada}.

\vspace{1cm}

For $k^{*2}\not=0$ one  defines a slope parameter of the $P\gamma\gamma^*$
transition form factor by
\begin{equation}
\left. b_P\equiv \frac{1}{A(P\to \gamma\gamma)} \frac{d}{d k^{*2}} 
A(P\to \gamma \gamma^*) \right|_{k^{*2}=0},
\end{equation}
related to the pseudoscalar structure or dimensions.
The tree-level amplitude (\ref{pgg}) predicts $b_P=0$.  
Experimentally, there is a single measurement in the time-like region, 
using $\eta \to \gamma \mu^+ \mu^-$
\cite{Lepton-G} leading to 
\begin{equation}
\label{tlike}
b_\eta= (1.9 \pm 0.4)~ GeV^{-2}. 
\end{equation}
In addition,
there are also several sets of measurements of the process $\gamma \gamma^* \to P$
in electron--positron collisions for rather large values of $-k^{*2}$. 
The most recent set, when  
extrapolated to small values using a pole approximation, finds \cite{cleo} 
\begin{equation}
\label{slike}
b_\eta=1.67 \pm 0.02~GeV^{-2}.
\end{equation}

The next to leading order correction
in ChPT was computed for on--shell--photons by 
Donoghue et al. \cite{Donoghue} and Bijnens et al. 
\cite{BBC} and the expected 
cancellation of the divergences was explicitly found. 
However, when one allows for (at least one) off--shell final
photon, $k^{*2}\not=0$, the cancellation no longer occurs. 
The amplitude reads \cite{BBC}
\begin{equation}
\begin{array}{l}
\sd A(P\to\gamma\gamma^*)= -\alpha \frac{\sqrt{2} C_P}{f_P} 
\epsilon^{\mu \nu \alpha \beta} \epsilon_\mu k_\nu \epsilon^*_\alpha
k^*_\beta        \\
\phantom{A(P\to\gamma\gamma^*)}
\sd\times \Bigg( 1 + \frac{1}{16 \pi^2 f^2}  \left\{ \frac{2}{3} 
\lambda k^{*2} 
-\frac{1}{3}k^{*2}(\ln \frac{m_K^2}{\mu^2} + \ln \frac{m_\pi^2}{\mu^2})\right. \\
\phantom{A(P\to\gamma\gamma^*)}
\sd+\left. \frac{10}{9} k^{*2}+
\frac{4}{3}[F(k^{*2},m_\pi^2) + F(k^{*2},m_K^2)] \right\} \Bigg),
\end{array}
\end{equation}
where the first factor corresponds to the tree-level contribution in terms
of physical $f_P$, $\lambda\equiv \frac{1}{\epsilon} + 1+ \ln 4\pi
-\gamma$ ($\gamma$ is the Euler constant) contains the divergent terms, and
\begin{equation}
\label{F}
F(m^2,x)\equiv m^2 \left(1 -\frac{x}{4} \right) \sqrt{\frac{x-4}{x}} \ln
\frac{\sqrt{x}+\sqrt{x-4}}{-\sqrt{x}+\sqrt{x-4}} -2 m^2, \qquad x \equiv
\frac{k^{*2}}{m^2}.
\end{equation}
For small $k^{*2}$, $F(m^2,k^{*2}) \simeq -\frac{2}{3} k^{*2}$ and, in the
case of real photons, $k^2=k^{*2}=0$, all the effects of loops and
non-anomalous ${\cal L}_4$ terms reduce to change $f$ into $f_P$ for
$P=\eta_8,\eta_1$. In order to give a finite result for virtual photons
one has to include  
counter-terms of $O(p^6)$ in the anomalous sector, which are of the type \footnote{
A complete analysis of the  $O(p^6)$ counter-terms in the anomalous 
sector can be found in
Refs.\cite{fearing,ISSLER,BIJNENS1}}
\begin{equation}
C \epsilon^{\mu\nu\alpha\beta}\partial^\lambda F_{\lambda\nu}
F_{\alpha\beta} \left( tr [Q^2\Sigma\partial_\mu\Sigma^\dagger
-Q^2\Sigma^\dagger\partial_\mu\Sigma]
+ tr [Q\Sigma Q \partial_\mu\Sigma^\dagger
-Q\Sigma^\dagger Q \partial_\mu\Sigma]\right).
\end{equation}
The analysis was done by Bijnens et al. \cite{BBC}
assuming saturation by vector mesons in a hidden symmetry formulation.
Using $\theta=-19.5^\circ$, the slope parameter reads \cite{ABBCbr} 
\begin{equation}
\label{beta}
\sd b_\eta= 
 [\frac{2 f_{\eta_1} + f_{\eta_8}}{2 f_{\eta_1} + 2 f_{\eta_8}}\frac{1}{16 \pi^2 f^2} \frac{-1}{3} (2 + \ln
\frac{m_\pi^2}{\mu^2} \frac{m_K^2}{\mu^2})] + \frac{1}{\mu^2} 
\end{equation}
where the term inside the bracket 
corresponds to the chiral loops and the other to the 
counter-terms.
Using $f_{\eta_8}=1.3 f_\pi, f_{\eta_1}=1.1 f_\pi$ and fixing the scale $\mu=M_\rho$, 
one obtains
$b_\eta \simeq [0.19 + 1.70]~ GeV^{-2} \simeq 1.89~ GeV^{-2}$,
being the major contribution due to the counter-terms.
The agreement with the experimental measurement 
Eq.(\ref{tlike}), performed in the time-like region, is very good. 
However, it differs substantially
from the more precise  measurement Eq.(\ref{slike}), obtained in the space-like
region.
It is amusing to observe that $SU(3)$ breaking effects can be considered
not only for the decay constants, but also for 
the renormalization scale $\mu$. In so doing, the relevant scale parameter 
is not $M_\rho$ but the mean vector-meson mass, defined through its squared as 
$\mu^2= (9 M_\rho^2 + M_\omega^2+2 M_\phi^2)/12 = 0.69~GeV^2$ \cite{ABBCbr}.
Using this scale in (\ref{beta}) one gets $b_\eta= 1.69~GeV^{-2}$, in
nice agreement with Eq.(\ref{slike}).


\subsection{$\bf \eta \to \mu^+\mu^-,~e^+ e^-$}

The $\eta$ decays into lepton pairs also are of e.m. origin. At present, 
data exist for the muon channel, with  
a branching ratio 
$BR(\eta \to \mu^+\mu^-)\equiv\Gamma(\eta \to \mu^+\mu^-)/\Gamma(\eta \to all)=
(5.8\pm 0.8) \times10^{-6}$ \cite{PDG}.
Normalizing this result to $\eta \to \gamma\gamma$, one gets
\begin{equation}
\label{etaexp}
B(\eta \to \mu^+\mu^-) \equiv {\Gamma(\eta \to \mu^+\mu^-) \over
\Gamma(\eta \to \gamma\gamma)} = (1.47 \pm 0.21) \times 10^{-5} \ , 
\end{equation}
where the branching ratio \cite{PDG} $BR(\eta \to \gamma\gamma) = 0.393\pm0.0025$
has been used. 

The ``reduced'' ratio (\ref{etaexp}) can be expressed in terms of a dimensionless
``reduced'' complex amplitude $R(\eta \to \mu^+\mu^-) $, normalized to the 
intermediate $\eta \to \gamma\gamma$ amplitude, leading to 
\begin{equation}
\label{bpll}
B(\eta \to \mu^+\mu^-) = 2\beta \left( {\alpha \over \pi} {m_l \over m_P}
\right)^2 |R(\eta \to \mu^+\mu^-)|^2,       
\end{equation}
where $\beta = \sqrt{1 - 4m_\mu^2/m_\eta^2}$. The on-shell $\gamma\gamma$ 
intermediate state generates the model independent imaginary part of $R$
\begin{equation}
{\rm Im}\, R(\eta \to \mu^+\mu^-) = {\pi \over 2 \beta} 
\ln{1-\beta \over 1+\beta}= -5.47~.  
\end{equation} 
The unitary bound on $B$, $B \geq B^{unit}$, is then obtained by setting 
${\rm Re}\, R = 0$ in Eq.(\ref{bpll}). It takes the value
\begin{equation}
B^{unit} (\eta \to \mu^+\mu^-)  =  1.11 \times 10^{-5}. 
\end{equation} 
In terms of $B^{unit}$, the PDG's result \cite{PDG} reads
${B(\eta \to \mu^+\mu^-) / B^{unit}}= 1 + (Re R/Im R)^2  =  1.3 \pm 0.2$
and can be used to extract ${\rm Re} R$ from experiment  
\begin{equation}
\label{realeta}
{\rm Re}\, R(\eta \to \mu^+\mu^-)   =  
\pm \left(3.0^{+0.9}_{-1.2} \right) .
\end{equation}

From the theoretical point of view and assuming the obvious dominance of
the two photon contribution, the reduced amplitude 
$R(q^2) = R(P \to l^+l^-)$ can be written as \cite{ABM}
\begin{equation}
\label{R}
R(q^2) = {2i \over \pi^2 q^2}
\int d^4k {q^2 k^2 - (q\cdot k)^2 \over k^2(q-k)^2[(p-k)^2-m_l^2]} 
F(k^2,k^{*2}),
\end{equation}
where $q^2=m_P^2,p^2=m_l^2$ and $k^*=q-k$, and $F$ is a generic and
model-dependent form factor, with $F(0,0)=1$ for on-shell photons.
The imaginary part of $R$ is finite and model-independent. By contrast,
its real part contains an {\it a priori} divergent $\gamma\gamma$ loop
(if a constant $F(k^2,k^{*2})=1$ form factor is assumed). The cure to
this problem is model dependent, and proceeds 
either through the inclusion of non-trivial 
form factors --which depend on the hadronic physics governing the
$P\to\gamma^*\gamma^*$ transition-- or, in a more modern ChPT language, the
inclusion of local counter-terms to render the result finite.

Predictions for several form-factor  models are
known \cite{pll}. Let us mention here the simple  Vector Meson Dominance  (VMD)
approach, which essentially implies neglecting direct $P\gamma\gamma$
and $VP\gamma$ vertices, thus assuming the full dominance of the chain
$P\to VV\to\gamma\gamma$. The corresponding form factor is
\begin{equation} 
\label{FVV} 
F=F_{VV}={M_V^2 \over M_V^2-k^2}{M_V^2
\over M_V^2-k'^2}, 
\end{equation}
which, when plugged in (\ref{R}), leads to 
\cite{QuiggJ,ABM},
\begin{equation}
{\rm Re}\, R (\eta \to \mu^+\mu^-)=
 -1.3^{+0.7}_{-0.5},
\end{equation} 
 in the $SU(3)$-symmetric limit $M_V=M_{\rho,\omega}$.
 This corresponds to 
 \begin{equation}
 \label{Bmmsu3}
  B(\eta \to \mu^+\mu^-) = 
(1.18^{+0.08}_{-0.06})\times 10^{-5},
\end{equation}
 in reasonable agreement with the experimental value 
(\ref{etaexp}).
For completeness, we also quote the 
corresponding results for the $\eta \to e^+e^-$ decay amplitude:
$\quad {\rm Re}\, R(\eta \to e^+e^-)=31.3\pm2.0, \quad
{\rm Im}\, R(\eta \to e^+e^-)=-21.9$\quad and 
$\quad B(\eta \to e^+e^-)=(3.04\pm0.26)B^{unit}=(1.37\pm0.12) \times 10^{-8}$.

From the ChPT point of view, Savage et al.~\cite{Wise} have introduced
local $P l^+l^-$ counter-terms to
render the amplitude finite, 
\begin{equation}
{\cal L}_{\rm c.t.}
= {3i\alpha^2\over 32\pi^2}
\overline{l}\gamma^\mu\gamma_5 l\
\left[\chi_1Tr(Q^2\Sigma^\dagger\partial_\mu \Sigma -
Q^2\partial_\mu\Sigma^\dagger\Sigma)\right.
\left. + \chi_2Tr(Q\Sigma^\dagger Q\partial_\mu\Sigma
-Q\partial_\mu\Sigma^\dagger Q\Sigma)\right]\ ,
\end{equation}
where $l=e$ or $\mu$, and $Q$ is the electromagnetic quark charge matrix.
Using the $\overline{MS}$ minimal subtraction scheme in dimensional
regularization, the ``reduced'' amplitude reads 
\cite{ABM}
\begin{equation}
\label{our}
\begin{array}{l}
{\rm Re}\, R(q^2=m_P^2)
= -\strut\displaystyle{\chi_1(\Lambda)+\chi_2(\Lambda)\over 4} - {5\over 2}+ 3 
\ln{m_l \over \Lambda} \\
~ \\
\phantom{Re R=(q^2=m_P^2) }
\strut\displaystyle + {1\over 4\beta}\ln^2 {1-\beta\over 1 + \beta}  
+ {\pi^2\over 12 \beta} + {1\over \beta}  
Li_2\left({\beta-1\over \beta+1}\right),
\end{array}
\end{equation}
where the explicit logarithmic dependence on $\Lambda$ reflects the 
ultraviolate divergence of 
the loop, and cancels with the inclusion of the local counter-terms 
$\chi_1(\Lambda)+\chi_2(\Lambda)$. 
The present experimental $\eta\to\mu^+\mu^-$ branching ratio
requires a counter-term (for $\Lambda = M_\rho=0.77$ GeV) given by    
\begin{equation}
\label{countrho}
\chi_1(M_\rho) + \chi_2(M_\rho) = \left\{ \begin{array}{r}  -7^{+4}_{-5} \\
                                                            -31^{+5}_{-4} 
\end{array} \right. .
\end{equation}
In turn, the less precise $\pi^0\to e^+ e^-$ available experimental data
\cite{pion} 
translate correspondingly into
\begin{equation}
\label{countrhopi}
\chi_1(M_\rho) + \chi_2(M_\rho) = \left\{ \begin{array}{r}  -22^{+25}_{-16} \\
                                                            +81^{+16}_{-25} 
\end{array} \right. .
\end{equation}
Notice that 
the first values are consistent with the existence of a unique
counter-term, assuming lepton universality, while the second ones are not and, 
consequently, have to be 
discarded. This strongly relates $\eta\to\mu^+\mu^-$ and 
$\pi^0\to e^+ e^-$ in the sense that an accurate  measurement  
for one of the two
processes allows a good prediction for the other.

Adopting the resonance saturation hypothesis, one can go one step 
further and predict the value of the finite part of these counter-terms. 
This amounts to choose $M_{\rho,\omega}~=~0.77$~GeV
 as both the subtraction point 
$\Lambda$ and the mass $M_V$ appearing in our $F_{VV}$ form factor
(\ref{FVV}).  The previous result for $\eta \to \mu^+ \mu^-$ (\ref{Bmmsu3})
can be taken also as the prediction from ChPT with resonance saturation.
In this context, it  leads to     
$\chi_1(M_\rho) + \chi_2(M_\rho) \simeq  -14 $, 
close to the experimental values displayed in the first row of 
(\ref{countrho},\ref{countrhopi}). An alternative method to estimate
the counter-terms has been proposed recently using Lowest Meson Dominance 
in large-$N_c$ QCD \cite{peris}, with the result 
$\chi_1(M_\rho) + \chi_2(M_\rho) \simeq   -8.8 \pm 3.6~ ,$
in nice agreement with the first value in Eq.(\ref{countrho}).

We conclude that both the experimental result (\ref{etaexp}) and 
reasonable models indicate
$B(\eta\to \mu^+ \mu^-)\simeq 10^{-5}$, which is well in the range of Wasa 
possibilities. The corresponding predictions for
$B(\eta\to e^+ e^-)\sim 10^{-8}$ indicate that this channel might still
be detected at Wasa.

\subsection[9]{{\bf $\eta \to \pi^+ \pi^- \gamma$ } \footnotemark} 
\footnotetext{Notice that
this decay is also analyzed in Holstein's contribution to this Handbook 
\cite{holstein}}

The lowest order amplitude in ChPT
for $\eta(p_0) \to \pi^+(p_1) \pi^-(p_2) \gamma(k)$ can be easily 
calculated from the WZ action (\ref{WZ}). One obtains 

\begin{equation}
  \label{TREEA}
   A = {- e C_P' \over \sqrt{2} \pi^2 f^3}
       \epsilon^{\mu \nu \alpha \beta} \epsilon_\mu p_{1_\nu}
                                       p_{2_\alpha} p_{0_\beta},
\end{equation}
where $C_{\eta_8}' =1/\sqrt{3} $,  $C_{\eta_1}'=\sqrt{2}/\sqrt{3}$
and $\epsilon$ is the photon polarization. 
The reliability of this approach can be tested observing that the 
related $\gamma \pi \pi \pi$ coupling can be expressed in terms 
of the coupling constant
$F^{3\pi}$ ($C_\pi'=1$ in Eq.(\ref{TREEA})), 
which at this lowest order is related to the $\pi^0 \gamma \gamma$
coupling, $F^\pi$ (see Eq.(\ref{pgg}), $C_{\pi^0}=1$), by 
\begin{equation}
    F^{3\pi} = {e \over \sqrt{2} \pi^2 f^3}
             = {2 F^\pi \over e f^2}.
\end{equation}
From the first equality one obtains $F^{3\pi} = 9.7 \pm 0.1\;
GeV^{-3},$ not far from the 
experimental value \cite{ANTIPOV}: $F^{3\pi}_{exp} = 12.9 \pm
0.9 \pm 0.5 \; GeV^{-3}$. The agreement is improved when the
extraction of the experimental value is corrected taking into 
account both higher
order chiral contributions \cite{hannah} and e.m. effects. One then
obtains  
$F^{3\pi}_{exp~corrected} = 10.7\pm 1.2 \; GeV^{-3}$ \cite{akt}.

The use of the tree-level
amplitude, Eq.(\ref{TREEA}), leads to $\Gamma(\eta\to
\pi^+ \pi^- \gamma)= 35 ~eV$, somewhat lower than the
experimental value, $\Gamma_{exp}(\eta\to
\pi^+ \pi^- \gamma)= 58 \pm 6 ~eV$ \cite{PDG},
and to a photon energy spectrum less
softer than the observed one \cite{LAYTER}. 
Notice that  use of $f_{\eta_8} =  f_\pi$ 
--as predicted in ChPT at this level--, has been made in the 
above calculation,  
as well as $f_{\eta_1} =  f_\pi$, assuming nonet symmetry.
Next to leading 
corrections break this $U(3)$ symmetry results and have been estimated to imply 
$f_{\eta_8}/f_\pi = 1.3$, $f_{\eta_1}/f_\pi=1.1$ \cite{BIJNENS1} as
discussed in subsection 2.1 where a good description of 
$\pi^0,\eta,\eta'\to\gamma\gamma$ was achieved using
$\theta=-19.5^\circ$. The introduction of 
the $SU(3)$ breaking in the decay constants requires one-loop contributions.

The one-loop diagrams contributing to the processes
$\eta,\eta^\prime \to \pi^+ \pi^- \gamma$
have been calculated using dimensional regularization in
Ref.\cite{BIJNENS2}.
The results can be conveniently written in terms of divergent parts,
containing $\lambda = 1/\epsilon + 1 +\ln 4 \pi - \gamma$, 
and the function $F$ in Eq.(\ref{F}). 
The amplitudes for the processes relevant for the $\eta$ particle turn out to be
\begin{equation}
   \label{LOOPA2}
      \begin{array}{rl}
 A(\eta_8 & \to \pi^+ \pi^- \gamma)  =
                 \sd  {- e \over \sqrt{6} \pi^2 f_\pi^2 f_{\eta_8}}
            \epsilon^{\mu \nu \alpha \beta} A_\mu p_{1_\nu}
                                       p_{2_\alpha} p_{0_\beta} \\
~~~~~ & ~~~~~~~~~~~~~~~~~~~~~~~~~~~~~~~~~~~~~~~~~~~~~~~~~~~~~~~  \\
~~~~~ & \sd \{ 1 + {1 \over 16 \pi^2 f^2} [\lambda
                    (4 m^2_\pi - 4 m^2_K + p^2_{12}+ k^2)
                -(4 m^2_\pi + {1 \over 3} p^2_{12}) \ln
                                            {m^2_\pi \over \mu^2}
             - k^2 \ln{m_K^2 \over \mu^2}
                                            \\
~~~~~ & ~~~~~~~~~~~~~~~~~~~~~~~~~~~~~~~~~~~~~~~~~~~~~~~~~~~~~~~  \\
~~~~~ &    \sd   +(4 m^2_K - {2 \over 3} p^2_{12}) \ln
                                            {m^2_K \over \mu^2}
                +{5 \over 3} p^2_{12} +{5\over 3} k^2
                +{4 \over 3} F(m_\pi^2,p^2_{12})
                +{8 \over 3} F(m_K^2,p^2_{12})]  \\
		~~ & ~~~ \\
~~~~~ &                 +4 F(m_K^2,k^2) 
                 \},
\end{array}
\end{equation}
\begin{equation}
   \label{LOOPA3}
      \begin{array}{rl}
 A(\eta_1 & \to \pi^+ \pi^- \gamma)  =
                \sd   {- e \over \sqrt{3} \pi^2 f_\pi^2 f_{\eta_1}}
            \epsilon^{\mu \nu \alpha \beta} A_\mu p_{1_\nu}
                                       p_{2_\alpha} p_{0_\beta} \\
~~~~~ & ~~~~~~~~~~~~~~~~~~~~~~~~~~~~~~~~~~~~~~~~~~~~~~~~~~~~~~~  \\
~~~~~ & \sd \{ 1 + {1 \over 16 \pi^2 f^2} [\lambda
                    (4 m^2_\pi + 2 m^2_K + {1 \over 2} p^2_{12})
                -(4 m^2_\pi + {1 \over 3} p^2_{12}) \ln
                                            {m^2_\pi \over \mu^2}\\
~~~~~ & ~~~~~~~~~~~~~~~~~~~~~~~~~~~~~~~~~~~~~~~~~~~~~~~~~~~~~~~  \\
~~~~~ &   \sd    -(2 m^2_K + {1 \over 6} p^2_{12}) \ln
                                            {m^2_K \over \mu^2}
                +{5 \over 6} p^2_{12}
                +{4 \over 3} F(m_\pi^2,p^2_{12})
                +{2 \over 3} F(m_K^2,p^2_{12})] \}.
     \end{array}
\end{equation}
valid for both real ($k^2=0$) and virtual ($k^2>0$) photons.

The divergences appearing in the loop calculation of the
amplitudes are eliminated redefining the coefficients of the
$O(p^6)$ Lagrangian. The terms contributing to the
$\eta_8 \pi^+ \pi^- \gamma$ vertex can
be easily read from the general expression obtained in Ref.
\cite{BIJNENS1,ISSLER,fearing}. The relevant terms are
\begin{equation}
   \label{P6TERMS}
     \begin{array}{ll}
  {\cal L}_6^\epsilon & = C_1 \epsilon^{\mu \nu \alpha \beta}  F_{\mu \nu}
            tr [Q (\partial_\alpha \Sigma^\dagger \Sigma
                  +\partial_\alpha \Sigma \Sigma^\dagger)]
            tr (\chi \partial_\beta \Sigma^\dagger
               +\chi^\dagger \partial_\beta \Sigma) \\
	       \ \\
 ~~         & + C_2 \epsilon^{\mu \nu \alpha \beta}  F_{\mu \nu}
                tr[(\Sigma Q \partial_\alpha \Sigma^\dagger
               +Q \Sigma \partial_\alpha \Sigma^\dagger
               -\partial_\alpha \Sigma Q \Sigma^\dagger
               +\Sigma \partial_\alpha \Sigma^\dagger Q) \\
		  \ \\
~~ &               (\partial_\lambda \partial_\beta \Sigma
                         \partial^\lambda \Sigma^\dagger
               -\partial_\lambda \Sigma
                        \partial_\beta \partial^\lambda \Sigma^\dagger
               +\partial_\beta \Sigma
                       \partial^2 \Sigma^\dagger
               -\partial^2 \Sigma
                     \partial_\beta \Sigma^\dagger) ]\\
		     \ \\
  &            + C_3 \epsilon^{\mu\nu\alpha\beta}\partial^{\lambda}
                 F_{\lambda\mu} tr [Q\Sigma \partial_{\nu}\Sigma^\dagger
\Sigma \partial_\alpha \Sigma^\dagger\Sigma \partial_\beta\Sigma^\dagger],
     \end{array}
\end{equation}
where the first and the second terms take care of the
divergences proportional to the pseudoscalar masses and
invariant momenta, respectively. The terms of the Lagrangian in
Eq. (\ref{P6TERMS}) are not enough to eliminate the divergences
for $\eta_1 \to \pi^+ \pi^- \gamma$. This is because the
calculation in Ref. \cite{BIJNENS1,ISSLER} was performed for
$SU(3)_L \times SU(3)_R$ broken to $SU(3)_V$. When extending
the symmetry to include the singlet, new terms such as
\begin{equation}
   \label{P6TERMS2}
 \epsilon^{\mu \nu \alpha \beta} A_{\mu \nu}
  tr [(\chi \Sigma^\dagger + \Sigma \chi^\dagger)
      (\Sigma Q \Sigma^\dagger - Q) \Sigma \partial_\alpha \Sigma^\dagger]
  tr (\Sigma \partial_\beta \Sigma^\dagger) ,
\end{equation}
appear contributing only to this singlet part.

Numerical predictions cannot be automatically
made at this point because the
finite part of the coefficients in ${\cal L}_6^\epsilon$ are unknown  
constants. Here we will assume again that at $\mu \sim M_\rho$ resonance
saturation works and we fix the value of the relevant
coefficients by the vector meson contributions \`a la Bando \cite{Bando}.
Then, the complete expressions for the amplitudes of the processes we
are studying are
\begin{equation}
   \label{ACOMPLETE}
      \begin{array}{ll}
 A(\eta_8 \to \pi^+ \pi^- \gamma) & = \sd
                   {- e \over \sqrt{6} \pi^2 f_\pi^2 f_{\eta_8}}
            \epsilon^{\mu \nu \alpha \beta} A_\mu p_{1_\nu}
                                       p_{2_\alpha} p_{0_\beta} 
                                        \{ 1 + C_{loops}^{\eta_8}
             + {3 \over 2 m_\rho^2} p^2_{12} \},  \\
~~~~~~~~ & ~~~~~~~~~~~~~~~~~~~~~~~~~~~~~~~~~~~~~~~~~~~~~~~~~~~~~~~~~~~~~\\
 A(\eta_1 \to \pi^+ \pi^- \gamma) & = \sd
                   {- e \over \sqrt{3} \pi^2 f_\pi^2 f_{\eta_1}}
            \epsilon^{\mu \nu \alpha \beta} A_\mu p_{1_\nu}
                                       p_{2_\alpha} p_{0_\beta} 
                                        \{ 1 + C_{loops}^{\eta_1}
             + {3 \over 2 m_\rho^2} p^2_{12} \},
    \end{array}
\end{equation}
where the last pieces come from resonance saturated counter-terms and
$C_{loops}^{\eta_8}$ and
$C_{loops}^{\eta_1}$ are the finite part 
of the loop corrections quoted in Eqs.(\ref{LOOPA2},\ref{LOOPA3}). The $k^2$
term in these Eqs. follows from the (phenomenologically
reasonable) assumption that there is no direct (contact) coupling between
vector--mesons and PPP--states and that the $P\gamma\gamma$ and $PV\gamma$
vertices proceed exclusively through $PVV$.

The predicted decay width for $\eta \to \pi^+ \pi^- \gamma$ at $O(p^6)$
in this ChPT approach
turns out to be $\Gamma(\eta \to \pi^+ \pi^- \gamma) = 47
\; eV$ \cite{BIJNENS2}, thus improving the tree-level prediction, but 
still $1.8\sigma$ too low compared to the experimental value.
The correction is now dominated by the contribution of the
$O(p^6)$ terms of the Lagrangian. Setting their coefficients to
be zero, the prediction for the decay width would be
$\Gamma_{loops} (\eta \to \pi^+ \pi^- \gamma) = 27 \pm 3 \; eV$,
lower than the tree-level prediction and very different from the
experimental value. 
Notice that the above results correspond to the $\eta-\eta'$ mixing
angle $\theta=-19.5^\circ$. 
If one allows small deviations, $\theta=(-19.5\mp 2.5)^\circ$, the decay width becomes
$\Gamma(\eta \to \pi^+ \pi^- \gamma) = 47 \pm 5$~eV.

Wasa could provide accurate values of this branching
ratio as well as for the corresponding photonic spectrum (details on the
later can be found in Ref.\cite{BIJNENS2}).

\section{Even intrinsic parity $\eta$ decays}

In this section we move to the consideration of the 
even intrinsic parity $\eta$ decays (sometimes called non-anomalous
processes).
We start with a brief discussion of the hadronic channel 
$\eta\to 3\pi$, which is 
an isospin violating process and contains information on $m_d-m_u$.
Then we move to e.m. decays. 
We consider the decay $\eta\to\pi^0\gamma\gamma$, which turns to be
very interesting from the ChPT point of view. The
related decay $\eta\to\pi^0 l^+ l^-$, important as a possible check
of $C$ and $CP$ single photon contributions, is also commented.
Finally, the channel $\eta \to \pi^+ \pi^- \pi^0\gamma$, closely
related to $\eta\to 3\pi$ via photon radiation, is analyzed.

\subsection{$\bf \eta \to 3 \pi$}
\label{eta3pi}
The decay $\eta\to 3\pi$ 
was originally discussed in ChPT by Gasser and Leutwyler \cite{GLeta}.  It is 
also 
extensively discussed in Gasser's
\cite{gasser}, Bijnens \cite{hans} and Holstein's \cite{holstein} contributions to this
Handbook. Here we include a brief review in order to fix notation 
that will be used below when discussing $\eta\to\pi^+\pi^-\pi^0\gamma$
and $\eta\to\pi\pi\gamma\gamma$ decays.
The $\eta\to 3\pi$ transition violates G--parity. This makes the 
amplitude proportional to $m_d-m_u$ and therefore vanishes in the limit
of exact
isospin symmetry. Using the ChPT Lagrangian (\ref{L2lluis}) one finds the
well known tree-level amplitude (CA result)
\begin{equation}
\label{CA}
A(\eta\to \pi^+(p_+)\pi^-(p_-)\pi^0(p_0))= 
\sd - \frac{B(m_d-m_u)}{3\sqrt{3} f^2}
 \left[1 + 3\frac{s-s_0}{m_\eta^2-m_\pi^2}\right],
\end{equation}
where $s=(p_++p_-)^2$, $s_0=(s+t+u)/3=m_\pi^2+m_\eta^2/3$ and
$B(m_d-m_u)/2=\Delta m_K^2$ is the QCD part of the kaonic 
squared mass difference (see Eqs.(\ref{B},\ref{TADPOLE})). 
Gasser and Leutwyler \cite{GLeta} used the Dashen's theorem \cite{DASHEN} 
to fix 
\begin{equation}
\label{DT}
\Delta m_K^2 \big|_{DT}= m_{K^0}^2 - m_{K^+}^2 - m_{\pi^0}^2 + m_{\pi^2}^2=
5.3\times 10^{-3}~GeV^2,
\end{equation}
which leads to the lowest order rate 
\begin{equation}
\label{tree}
\Gamma_{DT}^{tree}(\eta\to\pi^+\pi^-\pi^0)=66~eV,
\end{equation}
well below the experimental result \cite{PDG}
\begin{equation}
\label{exp}
\Gamma^{exp}(\eta\to\pi^+\pi^-\pi^0)=271\pm 26~eV.
\end{equation}
This prompted the same authors  \cite{GLeta} to introduce the one-loop
corrections
plus the corresponding counter-terms. The one-loop expressions
for the amplitudes read 
\begin{eqnarray}
\label{etappponeloop}
&&A(\eta\to \pi^+(p_+)\pi^-(p_-)\pi^0(p_0))= 
\sd - \frac{B(m_d-m_u)}{3\sqrt{3} f^2}
 \left[1 + 3\frac{s-s_0}{m_\eta^2-m_\pi^2} + U + V +W \right],\nonumber \\
&&A(\eta\to 3\pi^0) = A(\eta\to \pi^+\pi^-\pi^0)+
                      A(\eta\to \pi^-\pi^0\pi^+)+
		      A(\eta\to \pi^0\pi^+\pi^-),
\end{eqnarray}
where $U + V$ contains the contribution of loop diagrams and $O(p^6)$
counter-terms, while $1+W$ takes into account the $\eta-\eta'$ mixing effects
(and depends essentially on the $L_7^r$ counter-term).
For further reference we quote  the corresponding values at the center of the
Dalitz plot:  $U_0 + V_0=0.39 - 0.03 + 0.18 i$ and  $1 + W_0\simeq
1.15$. 
The one-loop effects in \cite{GLeta} represent an enhancement by a factor
$2.4$ respect to the lowest order prediction 
\begin{equation}
\label{1loop}
\Gamma_{DT}^{1~loop}(\eta\to\pi^+\pi^-\pi^0)=160\pm 50~eV,
\end{equation}
where the error accounts for higher order uncertainties. This result is 
in qualitative agreement with the analysis 
done in Ref.\cite{Truong} where it is shown that 
the unitarity corrections generated by $\pi
\pi$ final state interactions are large. On the other hand, 
the inclusion of e.m. 
corrections do not modify appreciably the above result \cite{kambor}.

This result can be reanalyzed having in mind 
that Dashen's theorem
should be corrected \cite{DTnew} resulting in $\Delta m_K^2 \simeq 1.3 \Delta
m_K^2 \big|_{DT}$, which represents enhancements of $\sim 70\%$ 
to the previous values, thus going in the good direction to explain the
experimental rate. 

On the other hand, some discrepancies with the 
ChPT predictions seem to appear when
expanding the decay amplitude around the center of the Dalitz plot.
Moreover, when comparing the decay rates into neutral and charged pions, 
the situation is also confusing. Indeed,
in terms of the branching ratio
\begin{equation}
\label{ratio}
r=\sd\frac{\Gamma(\eta\to3\pi^0)}{\Gamma(\eta\to \pi^+\pi^-\pi^0)},
\end{equation}
one gets $r_{tree}=1.51$, $r_{1~loop}=1.43$ \cite{GLeta}, whereas the
experimental value coming from the PDG is $r_{exp}=1.40\pm 0.03$ \cite{PDG},
favoring the one-loop result,
while a recent measurement at the CMD-2 detector reports
\cite{Akhmetshin} 
$r=1.52\pm 0.04\pm 0.08$, which seems to favor the tree-level result,
although being compatible with the one-loop prediction.

\subsection{$\bf \eta \to \pi^0 \gamma\gamma$}

The $\eta \to \pi^0 \gamma\gamma$ decay has a long story, nicely reviewed
in Ref.\cite{achasov}, and furnish a stringent test of ChPT.
It is also discussed by Bijnens \cite{hans}.
At lowest order in ChPT, one predicts a vanishing $\eta\to \pi^0 \gamma\gamma$
decay width contrasting with the measured 
value \cite{PDG} \footnote{A recent measurement by Crystal Ball of 
$\Gamma(\eta\to\pi^0\gamma\gamma)=0.38\pm0.11$~eV 
was reported by Nefkens \cite{nefkens} during this workshop.}
\begin{equation}
\label{WIDTH}
\Gamma(\eta\to\pi^0\gamma\gamma)=(7.1\pm1.4)\times 10^{-4}\ \Gamma(\eta\to 
\hbox{all})=0.84\pm0.18~\hbox{eV.}
\end{equation}
Higher orders in ChPT allow to obtain predictions not 
far from this value thanks to a rather complicated interplay among loop 
corrections and counter\-terms.

The amplitude for the decay $\eta(P) \to \pi^0 (p) \gamma(q_1) \gamma(q_2)$
can be written in the following form
\begin{equation}
\label{MAT}
M= \sum_i (a_i^t+a_i^P)A + \sum_i (b_i^t+b_i^P) B,
\end{equation}
where $A$ and $B$ are the two kinematically allowed amplitudes
\begin{equation}
\label{AB}
\begin{array}{l}
A= (\epsilon_1\cdot\epsilon_2) (q_1\cdot q_2) - (\epsilon_1\cdot q_2) 
(\epsilon_2\cdot q_1) \\
~\\
B= -(\epsilon_1\cdot\epsilon_2) (P\cdot q_1) (P\cdot q_2)
-(\epsilon_1\cdot P) (\epsilon_2 \cdot P) (q_1\cdot q_2) \\
\qquad + (\epsilon_1\cdot q_2) (\epsilon_2 \cdot P) (P\cdot q_1)
+ (\epsilon_1\cdot P) (\epsilon_2 \cdot q_1) (P\cdot q_2).
\end{array}
\end{equation}
Their coefficients $a_i^t,~b_i^t$ refer to the tree--level and counter--term
contributions, whereas $a_i^P,~b_i^P$, with $P=\pi$ or $K$, refer to the 
contributions from pion or kaon loops. The sum extends from the lowest order 
terms, $i=2$, to higher ones $i=4,6,...$ We also define $s=(q_1+q_2)^2= 
2 q_1\cdot q_2$, $t=(P-q_2)^2$, $u=(P-q_1)^2$ and $s+t+u=m_\eta^2+ 
m_\pi^2$. With these conventions fixed we now proceed to review the 
different contributions to the amplitude $M$. 

As previously stated, there is no lowest order $O(p^2)$ contribution 
from ${\cal L}_2$ to the $\eta 
\to \pi^0 \gamma\gamma$ amplitude, i.e. $a_2^t=b_2^t=0$, as can be immediately 
seen from Eqs.(\ref{L2lluis}) and 
(\ref{COVDER}) with $[Q,\Sigma]=0$ for neutral pseudoscalars. 

At next order, $O(p^4)$, the tree--level amplitude coming from
Eq.(\ref{L4}) is also zero, $a_4^t=b_4^t=0$. This is clearly the case for 
the WZ--term contributing only to anomalous processes and also for the  
counter-terms in Eq.(\ref{L4}). 
Indeed, photon couplings are generated either by covariant 
derivatives or by $F_{\mu\nu}$ terms like the two terms 
proportional to  $L_9$ 
(for one photon) and  to $L_{10}$ (for two photons) in (\ref{L4}). 
But the vanishing of $[Q,\Sigma]$ 
when only neutral $\pi^0$ and $\eta$ are involved is enough to guarantee a 
null contribution in all cases \cite{BIJCOR}. This absence of counter-terms implies 
that loop contributions at this order cannot be divergent. They contain 
two vertices of the Lagrangian ${\cal L}_2$ and a (charged) pion or kaon loop. 
One obtains $b_4^{\pi,K}=0$ and the finite expressions \cite{ABBC}
\begin{equation}
\label{OP4}
\begin{array}{l}
a_4^{\pi}= \sd{-4\sqrt{2} \alpha \over 3\sqrt{3}\pi f^2} \Delta m_K^2 \left(1+ 
\sd{3s - m_\eta^2-3m_\pi^2 \over m_\eta^2-m_\pi^2}\right) H(s,m_\pi^2) \cr
~\\
a_4^{K}= \sd{2\sqrt{2} \alpha \over 3\sqrt{3}\pi f^2}  
\left(3s - m_\eta^2-{1\over 3}m_\pi^2 -{8 \over3} m_K^2\right) H(s,m_K^2) 
\end{array}
\end{equation}
with ($x\equiv s/m^2$)
\begin{equation}
\label{H}
\begin{array}{l}
s H(s,m^2)\equiv s \sd \int_0^1 dz \sd \int_0^{1-z} dy \sd{zy\over m^2-s zy}=
-\frac{1}{2}-\frac{1}{2 x}\ln^2 \frac{\sqrt{x} + \sqrt{x-4}}
{-\sqrt{x} + \sqrt{x-4}} \cr
~\\
sH(s,m_K^2)\simeq \sd{s\over 24 m_K^2}\qquad \hbox{for\ } s<< m_K^2. 
\end{array}
\end{equation}
The small correction to $a_4^K$ proportional to $\Delta m_K^2$ has been 
ignored.

If one further introduces the numerical values 
$f=130.7~\hbox{MeV}$, $\Delta m_K^2$ as in Eq.(\ref{TADPOLE}) and the $\eta$
mass, one obtains the loop contributions up to  $O(p^4)$ to
the $\eta \to \pi^0\gamma\gamma$ width
\begin{equation}
\label{GAMMAOP4}
\Gamma^{(4)} (\eta \to \pi^0\gamma\gamma)= 7.18\times 10^{-3}~\hbox{eV}
\end{equation}
more than two orders of magnitude below the measured width (\ref{WIDTH}).
The pion loop contribution is small due to the G--parity 
violation in the vertex $\eta\pi^+\pi^-\pi^0$, which makes the amplitude 
proportional to $m_d-m_u$ or $\Delta m_K^2$ as seen in Eq.(\ref{OP4}). 
The contribution of the G--parity conserving kaon loops is 
also small because the loop--integration introduces the function 
$H(s,m_K^2) \simeq 1/24 m_K^2$  with a large denominator in our case with 
$m_K^2>>s$. Further contributions are required. This was noted independently
by Ecker et al., Ref.\cite{ECKERRING}.

Corrections of $O(p^6)$ were partially considered in \cite{ABBC,ko} in terms
of resonance saturation for counter-terms including vectors as well as the
scalar $a_0(980)$ and tensor $a_2(1320)$ resonances. 
Such tree-level contributions
were argued to dominate over one-loop corrections of the same order
using G-parity suppression 
arguments in the case of pion loops, and the smallness
of the one-loop functions appearing in kaon loops. The corresponding
result was
\begin{equation}
\Gamma^{(6)}(\eta\to\pi^0\gamma\gamma)\simeq 0.18.
~\hbox{eV}
\end{equation}

At $O(p^8)$ more counter-terms appear and a new type of 
loop--correction becomes potentially important. Since the counter-term 
contributions are assumed to be resonance dominated one can simply take the 
full VMD amplitude,
\begin{equation}
\label{MVMD}
M_{VMD}=\sd{2\sqrt{2}\over 3\sqrt{3}} g_{\omega 
\pi^0\gamma}^2\left\{\Big[
\sd{P\cdot q_2-m_\eta^2\over M_V^2-t}+ \sd{P\cdot q_1-m_\eta^2\over 
M_V^2-u}\Big] A - \Big[\sd{1\over M_V^2-t} + \sd{1 \over M_V^2-u}\Big]
B \right\},
\end{equation}
as an ``all--order'' estimate, i.e., at 
$O(p^6)$ and higher. This represents a substantial increase to 
the rate, namely,
\begin{equation}
\label{VMD}
\Gamma_{VMD}(\eta\to\pi^0\gamma\gamma)= 0.31~\hbox{eV},
\end{equation}
in reasonable agreement with older VMD estimates \cite{VMDOLD}. Analogous 
$a_0(980)$ and $a_2(1320)$ contributions are expected to slightly modify this result 
as discussed above. A new type of loop effects looks {\sl a priori} more 
interesting. Taking two vertices from the anomalous ${L}_{WZ}$ implies
a non--anomalous one--loop correction of $O(p^8)$. Indeed, pion loops are no 
longer suppressed because there is no G--parity violation in the 
$\pi\pi\pi\gamma$ and $\eta\pi\pi\gamma$ vertices and 
moreover the kaon loop suppression due to the function $H$ in Eq.(\ref{H}) 
does not occur. However, one easily obtains
\begin{equation}
\label{G8}
\Gamma^{(8)}(\eta\to\pi^0\gamma\gamma)=4.8\times 10^{-3}~\hbox{eV} 
\end{equation}
to be compared with $\Gamma^{(4)}$ in Eq.(\ref{GAMMAOP4}). The 
dominant contribution to $\Gamma^{(8)}$ comes from the absorptive part 
of the amplitude, which is much larger than that of the pion loop in 
$\Gamma^{(4)}$ as expected. 

A convenient way of presenting the ChPT predictions for 
$\Gamma(\eta\to\pi^0\gamma\gamma)$ consists in summing up all the 
contributions which are not negligible and present no sign ambiguities, 
i.e., the non--anomalous pion and kaon loops at $O(p^4)$, 
Eq.(\ref{OP4}), the corresponding loops at $O(p^8)$ with two WZ 
vertices, and the ``all--order'' sum of VMD counter-terms in 
Eq.(\ref{MVMD}). This implies \cite{ABBC}
\begin{equation}
\label{GChPT}
\Gamma(\eta\to\pi^0\gamma\gamma)=0.42\pm 0.20~\hbox{eV},
\end{equation}
where the error comes essentially from the sign ambiguity 
in the interference  of the 
$a_0(980)$ and $a_2(1320)$ resonances respect to the
vector dominated contribution, and the lack of knowledge
of higher order contributions. 
This ChPT
result is in agreement with related work
by Picciotto \cite{Picciotto} (performed also in the same ChPT context)
and Ng and Peters \cite{Ng} (who estimated the dominant VMD contribution).
More phenomenological predictions using constituent quark models exist.
Good agreement with the experimental result can be reached with
constituent quark masses of $300~MeV$ for up and down quarks \cite{Ng2}.
Measures of the photon energy and/or the two-photon invariant
mass spectra, which have different shapes for different models, would help to  
elucidate the best theoretical description for $\eta\to
\pi^0\gamma\gamma$ decay.

\subsection{$\bf \eta \to \pi^0 l^+ l^-$}

The rare decays $\eta \to \pi^0 e^+ e^-, \pi^0 \mu^+ \mu^-$ have not yet 
been observed. They can occur at second order of the e.m.
interaction without $C$ violation, through  the chain $\eta \to \pi^0
\gamma \gamma \to \pi^0 e^+ e^-$ and $\pi^0\to\mu^+\mu^-$. 
The imaginary part of the amplitude is
given by forcing the two intermediate photons to be on--shell and requires
good knowledge of the $\eta \pi^0 \gamma\gamma$ decay which unfortunately
is not the case at the moment, as we discussed in the preceding subsection.
In addition, the real part of the amplitude diverges unless one introduces
suitable form factors for the off--shell photons that regularizes the loop
integration at expenses of making it model dependent. 

Some years ago,
Cheng \cite{Cheng} computed the $\eta \to \pi^0 e^+ e^-$ decay in the limit
$m_e=0$ using a VMD model. He found $\Gamma( \eta \to \pi^0 e^+ e^-)=
13~\mu eV$. More recently, Ng and Peters \cite{Ng} have analyzed the $\eta \to
\pi^0 \mu^+ \mu^-$ channel, where the muon mass can not be neglected.
Making also use of VMD (and also considering the $a_0(980)$ exchange in the two
photon intermediate state) they give approximate analytical expressions
for the imaginary part of the $\eta\to \pi^0 l \bar l$ amplitudes and for
the energy spectrum of the pion. The following unitarity bounds were obtained
\begin{equation}
\begin{array}{l}
\left. \Gamma^{unit}_{2\gamma}(\eta\to \pi^0 e^+ e^-) \right|_{VMD}=
1.1^{+0.6}_{-0.5}~ \mu eV \\
\left. \Gamma^{unit}_{2\gamma}(\eta\to \pi^0 \mu^+ \mu^-) \right|_{VMD}=
0.5^{+0.3}_{-0.2}~ \mu eV.
\end{array}
\end{equation}
The inclusion of the $a_0(980)$ exchange affects only the muon channel,
leading to $0.9^{+0.6}_{-0.5}$ ($0.3^{+0.4}_{-0.2}$) $\mu eV$ for
constructive (destructive) interference. They estimated the dispersive part
of the amplitude by means of an {\it ad hoc} cutoff of $1~GeV$ to be
$\left|Re A_{\mu\bar\mu}/ Im A_{\mu\bar\mu} \right| \le 0.5$, suggesting
that the rate can not be much greater than the unitarity bound. This conclusion
is consistent with the results obtained by Heiliger and Sehgal \cite{Sehgal}
using a dispersive treatment for the similar processes $K_L\to \pi^0 e^+ e^-$ 
and $K_l\to \pi^0 \mu^+ \mu^-$,  
$\left|Re A^{K_L}_{\mu\bar\mu}/ Im A^{K_L}_{\mu\bar\mu} \right| \le 0.7$,
$\left|Re A^{K_L}_{e\bar e}/ Im A^{K_L}_{e\bar e} \right| \le 1.25$.

\vskip0.5truecm
                   
\subsection{$\eta\to\pi^+\pi^-\pi^0\gamma$}
\label{eta3pigamma}
The $\eta\to\pi^+\pi^-\pi^0\gamma$ process has been analyzed in the 
ChPT framework \cite{peter,ambrosio}. At lowest order, the amplitude is
dominated by pure bremsstrahlung and is thus 
closely related to the isospin violating 
$\eta\to\pi^+\pi^-\pi^0$ amplitude. At $O(p^4)$, apart from the correction to the inner 
bremsstrahlung contribution, a new structure dependent (or direct
emission) contribution appears, which does not vanish in the isospin limit.
This makes the process interesting since it carries new information not
accessible from $\eta\to 3\pi$.
Both contributions were estimated in \cite{peter} using an {\it ad hoc}
parametrization of the $\eta\to\pi^+\pi^-\pi^0$ amplitude.
It essentially consists in
approximating the amplitude by its value at the center of the Dalitz
plot, using $U+V \simeq U_0+V_0 = 0.39 -0.03 + 0.18 i$; $1+W\simeq
\sqrt{2}$, value that differs from $1+W_0 \simeq 1.15$ and takes into
account the phenomenological $\eta-\eta'$ mixing for radiative processes, and 
using the enhancement of the amplitude through the correction of
Dashen's theorem.
This analysis was performed
considering only the isospin conserving kaon loop contribution, 
which was expected to be dominant. Next order contributions were also
estimated using VMD saturation hypothesis, in a similar way as
in the $\eta\to\pi^0\gamma\gamma$ ChPT prediction. The 
conclusion reached was that the inner bremsstrahlung
amplitude strongly dominates.
Several technical points were refined and improved in \cite{ambrosio}.
Mainly, a better treatment of the Low's theorem for the inner bremsstrahlung
amplitude and the inclusion of the pion-loops leads to the result
$BR(\eta\to\pi^+\pi^-\pi^0\gamma; E_\gamma \ge 10 MeV) = 
(3.14 \pm 0.05)\times 10^{-3}BR(\eta\to\pi^+\pi^-\pi^0)$.
The contribution of the structure dependent contribution to the above
result turns out to be a tiny $\sim 0.02\%$.

\section{$\eta\to\pi\pi\gamma\gamma$  decays}  

In this section we comment on the 
anomalous $\eta\to\pi\pi\gamma\gamma$ decays. 
They were first discussed
at tree-level in ChPT in Ref.\cite{scherer}. The next order
contributions to the neutral channel were computed in \cite{BI,ABBT}.
Being  multi-body decays, several contributions appear. In fact, the
amplitude  for the neutral channel, for instance, can be written as
\cite{ABBT}
\begin{eqnarray}
A(\eta\to\pi^0\pi^0\gamma\gamma)&=&
A(\eta\to\pi^0\pi^0\gamma\gamma)_{\pi^0-pole} \\
&+& A(\eta\to\pi^0\pi^0\gamma\gamma)_{\eta-tail}+
A(\eta\to\pi^0\pi^0\gamma\gamma)_{1PI}\\
&+& A(\eta\to\pi^0\pi^0\gamma\gamma)_{VMD}.
\end{eqnarray}
The first term proceeds through the chain 
$\eta\to\pi^0\pi^0\pi^0$ with  
a subsequent anomalous $\pi^0\to\gamma\gamma$ transition. It thus depends
on the non-anomalous and isospin breaking $\eta\to 3\pi^0$ amplitude, 
discussed in Section 
\ref{eta3pi}. Actually, the avaliable phase space suggests that a
simplified treatment of the $\eta\to 3\pi^0$, along the
lines described in the previous Section \ref{eta3pigamma} is possible. 
The second contribution proceeds through the isospin conserving decay
chain $\eta\to\pi^0\pi^0\eta\to\pi^0\pi^0\gamma\gamma$ and turns to
vanish in the chiral limit.
The third term includes one-particle irreducible (1PI) one-loop
diagrams, with an $\eta\to\pi^+\pi^-\gamma(\gamma)$ vertex from the WZ
Lagrangian, followed by $\pi^+\pi^- (\gamma) \to \pi^0\pi^0\gamma$
rescattering. The last VMD contribution represents an estimate of
higher-order contributions. The analysis of reference \cite{ABBT}
can be summarized in the diphoton mass spectrum of Fig.1. This figure
clearly shows that this partial decay width is dominated by the
pion-pole in $\eta\to 3\pi^0$.

In \cite{BI,ABBT} the charged $\eta\to\pi^+\pi^-\gamma\gamma$
channel was also discussed up to $O(p^6)$. The interest of this process from
the ChPT point of view is masked by the dynamics of the processes
$\eta\to\pi^+\pi^-\gamma$ with an additional radiative photon, and 
$\eta\to\pi^+\pi^-\pi^0$ with a subsequent $\pi^0\to\gamma\gamma$ decay.

\begin{figure}[t]
\epsfig{file=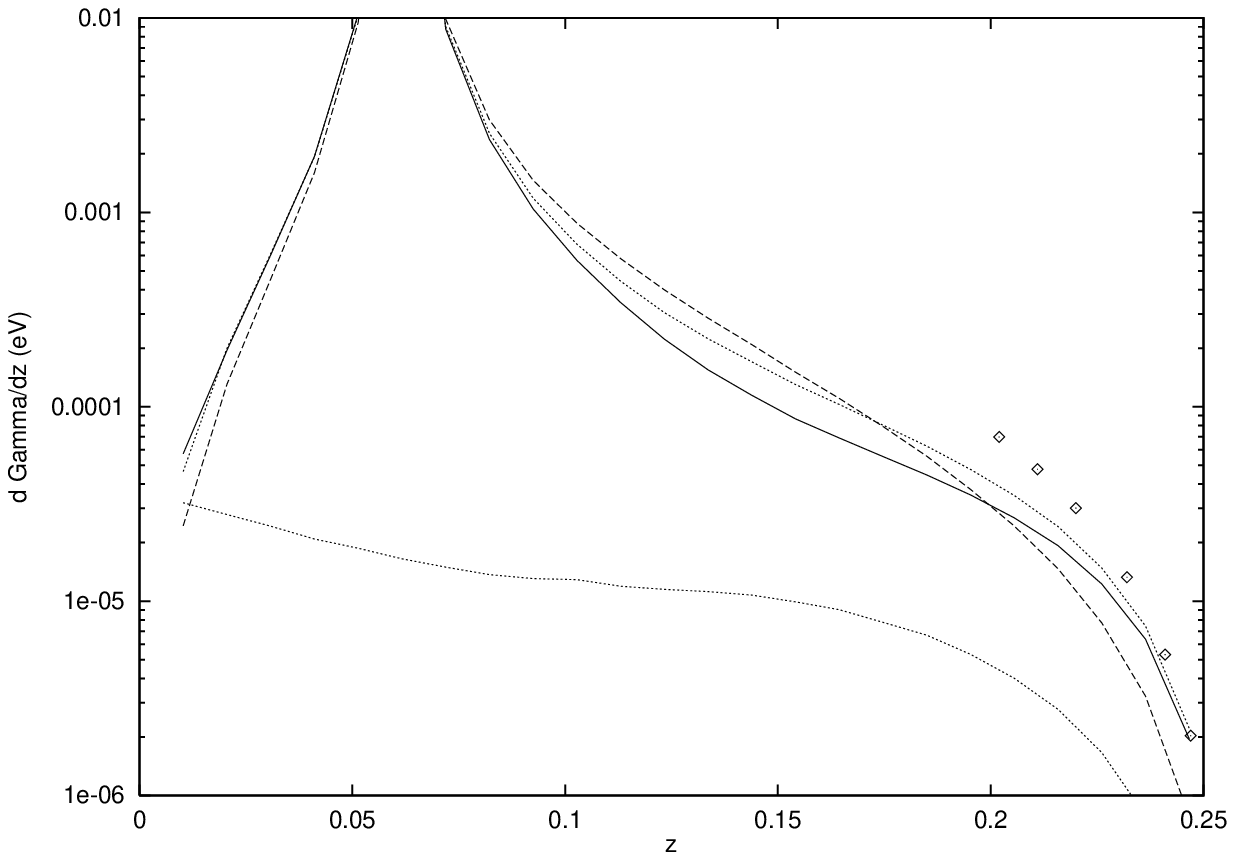,width=15cm,height=18cm,angle=0}
\caption[Diphoton mass spectrum $z\equiv m_{\gamma\gamma}^2/m_\eta^2 $ for the $\eta\to\pi^0\pi^0\gamma\gamma$
decay.]
{Diphoton mass spectrum  $z\equiv m_{\gamma\gamma}^2/m_\eta^2 $   for the $\eta\to\pi^0\pi^0\gamma\gamma$ decay.
The dashed line is  the $\pi^0$-pole contribution. The upper dotted
line corresponds to the $O(p^6)$ result. The lower dotted line is the VMD
contribution.
The solid line corresponds to the total contribution. 
The symbols at the end of the spectrum show the
total result using an approximate expression valid at values close to
$z =1/4$.}     
\end{figure}

\vskip1truecm

\noindent{\bf Acknowledgments.} 

I thank G. F\"aldt and J. Bijnens for their kind invitation to the 
Workshop on Eta Physics. I am also indebted to 
A. Bramon for his collaboration and careful
reading of the manuscript.

\vskip1.5truecm

\listoffigures

\end{document}